\documentclass[osajnl2,preprint,showpacs,superscriptaddress]{revtex4}  

\begin{document}

\title{Role of intensity fluctuations in third-order correlation \\ double-slit interference of thermal light}

\author{Xi-Hao Chen}
\affiliation{College of Physics, Liaoning University, Shenyang 110036, China}

\affiliation{Laboratory of Space Science Experiment Technology, Center for Space Science and Applied Research,
\\Chinese Academy of Sciences, Beijing 100190, China}

\author{Wen Chen}
\author{Shao-Ying Meng}
\author{Wei Wu}\email{wuweio@yahoo.com.cn}
\affiliation{College of Physics, Liaoning University, Shenyang 110036, China}

\author{Guang-Jie Zhai}
\affiliation{Laboratory of Space Science Experiment Technology, Center for Space Science and Applied Research,
\\Chinese Academy of Sciences, Beijing 100190, China}
\author{Ling-An Wu}
\affiliation{Laboratory of Optical Physics, Institute of Physics and Beijing National Laboratory for Condensed
Matter Physics,\\ Chinese Academy of Sciences, Beijing 100190, China}

\begin{abstract}
A third-order double-slit interference experiment with pseudo-thermal light source in the high-intensity limit has been performed by actually recording the intensities in three optical paths. It is shown that not only can the visibility be dramatically enhanced compared to the second-order case as previously theoretically predicted and shown experimentally, but also that the higher visibility is a consequence of the contribution of third-order correlation interaction terms, which is equal to the sum of all contributions from second-order correlation. It is interesting that, when the two reference detectors are scanned in opposite directions, negative values for the third-order correlation term of the intensity fluctuations may appear. The phenomenon can be completely explained by the theory of classical statistical optics, and is the first concrete demonstration of the influence of the third-order correlation terms.
\end{abstract}


\maketitle

The first optical intensity correlation experiment performed by Hanbury Brown and Twiss (HBT)in 1956~\cite{HBT} attracted extreme attention and even disbelief at the time. Similarly, the first ``ghost" interference (GI) experiment with two-photon entangled light~\cite{Strekalov}in 1995, followed some years later by second-order GI and ghost imaging with classical thermal light has given rise to wide interest and much warm discussion~\cite{Gatti,Cheng,lugiato,Zhu,wang,wang1,Valencia,Zhangda,Zhai,Zhai1,Scarcelli1,Shapiro,Chen}. Unfortunately, the drawback of second-order GI with thermal light is that the visibility
theoretically can never exceed 1/3. However, recent studies on the higher-order intensity correlation effects ~\cite{Liu,Liu1,Agafonov,Cao,Richter,Li,Agafonov1,Chen1,Boyd,Caoz}, show that by increasing the order $N$ the visibility rapidly increases and even approaches 1.

Many theoretical schemes for third and higher order GI with thermal light have been proposed~\cite{Liu,Li,Boyd,Caoz,Cao}, but due to the technical difficulties, few experimental demonstrations of high-order GI have been performed. The first multi-photon interference experiment with classical light was performed by Agafonov \textit{et al }~\cite{Agafonov}, who also proposed a high-order scheme to obtain improved visibility and signal-to-noise ratio for ghost imaging~\cite{Agafonov1}. The first $N$th-order ($N\geq2$) ghost imaging experiment~\cite{Chen1} in a lensless scheme was performed by recording only the intensities in two optical paths, which agreed well with theoretical predictions. Shih's group investigated the high-order coherence of thermal light based on Glauber's quantum optical coherence~\cite{Liuj} and then performed third-order HBT and ghost imaging experiments in the photon counting regime~\cite{Zhou,Zhou1}.

\begin{figure}[b]
\centerline{\includegraphics[width=15cm]{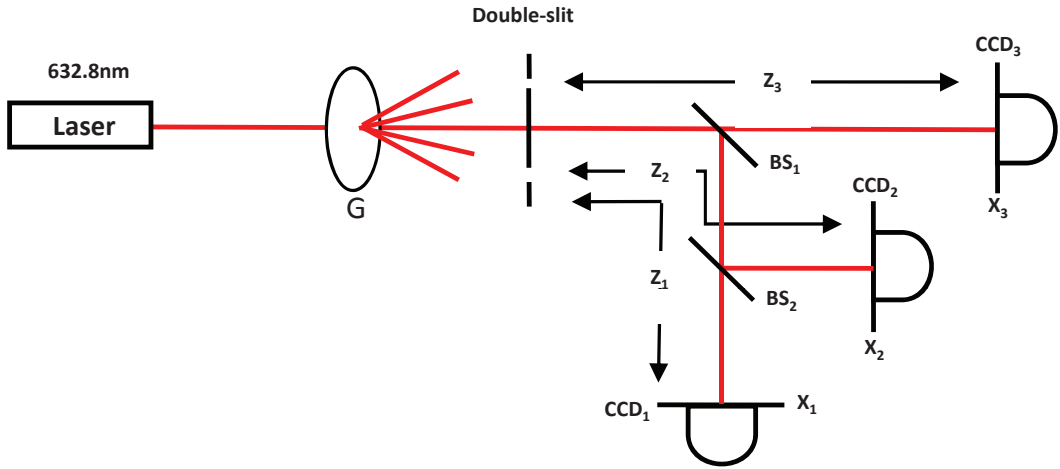}}\caption{(Color online)
Experimental setup for third-order double-slit interference with a
pseudo-thermal light source. G: rotating ground glass plate; BS:
beamsplitter; CCD$_i(i=1,2,3)$: CCD camera.} \label{scheme1}
\end{figure}

In this letter, we report the first (to our knowledge) demonstration of
third-order double-slit interference with pseudo-thermal light by recording the
intensities in three independent optical paths by means of three CCD cameras.
It is shown explicitly that the enhanced high visibility of third-order
interference is due to the contribution of the third-order intensity
fluctuation correlation, which equals the sum of all the contributions from the
second-order correlations, as previously pointed out by Liu et al \cite{Liu}
but never demonstrated experimentally before.

In the high-intensity
limit, the normalized second- and third-order correlation functions can be given by~\cite{Mandel}
\begin{equation} \label{g2}
g^{(2)}(x_1,x_2) =\frac{\langle I_1(x_1)I_2(x_2)\rangle}
 {\langle I_1(x_1)\rangle\langle I_2(x_2)\rangle},
\end{equation}
and
\begin{equation} \label{g3}
g^{(3)}(x_1,x_2,x_3) =\frac{\langle I_1(x_1)I_2(x_2)I_3(x_3)\rangle}
 {\langle I_1(x_1)\rangle\langle I_2(x_2)\rangle\langle I_3(x_3)\rangle},
\end{equation}
where $I_j(x_j)$($j=1,2,3$) is the instantaneous intensity at the
transverse position $x_j$ and $\langle...\rangle$ stands for ensemble
averaging.

According to ~\cite{Liu}, we can rewrite these normalized correlation functions for thermal light in terms of the intensity fluctuation correlation
\begin{equation}
g^{(2)}(x_1,x_2) =1+\Delta g^{(2)}_{12}(x_1,x_2)
\end{equation}
and

\begin{eqnarray}\label{sumg3}
\nonumber g^{(3)}(x_1,x_2,x_3) =1+\Delta g^{(2)}_{12}(x_1,x_2)+\Delta g^{(2)}_{23}(x_2,x_3)\\ +   \Delta g^{(2)}_{13}(x_1,x_3)+\Delta g^{(3)}_{123}(x_1,x_2,x_3)
\end{eqnarray}
where
\begin{equation}\label{deltg3}
\Delta g^{(3)}_{123}(x_1,x_2,x_3)=\frac{\langle  \Delta I_1(x_1) \Delta I_2(x_2)\Delta I_3(x_3)\rangle}{\langle I_1(x_1)\rangle\langle I_2(x_2)\rangle\langle I_3(x_3)\rangle}
\end{equation}
and
\begin{eqnarray}\label{deltg2}
\Delta g^{(2)}_{ij}(x_i,x_j)=\frac{\langle  \Delta I_i(x_i) \Delta I_j(x_j)\rangle}{\langle I_i(x_j)\rangle\langle I_i(x_j)\rangle}\\ \nonumber(i\neq j,\ i=1,2 \ and \ j=2,3 ),
\end{eqnarray}
in which  $\Delta I(x)=I(x)-\langle I(x)\rangle$.

\begin{figure}[t]
\centerline{\includegraphics[width=14cm]{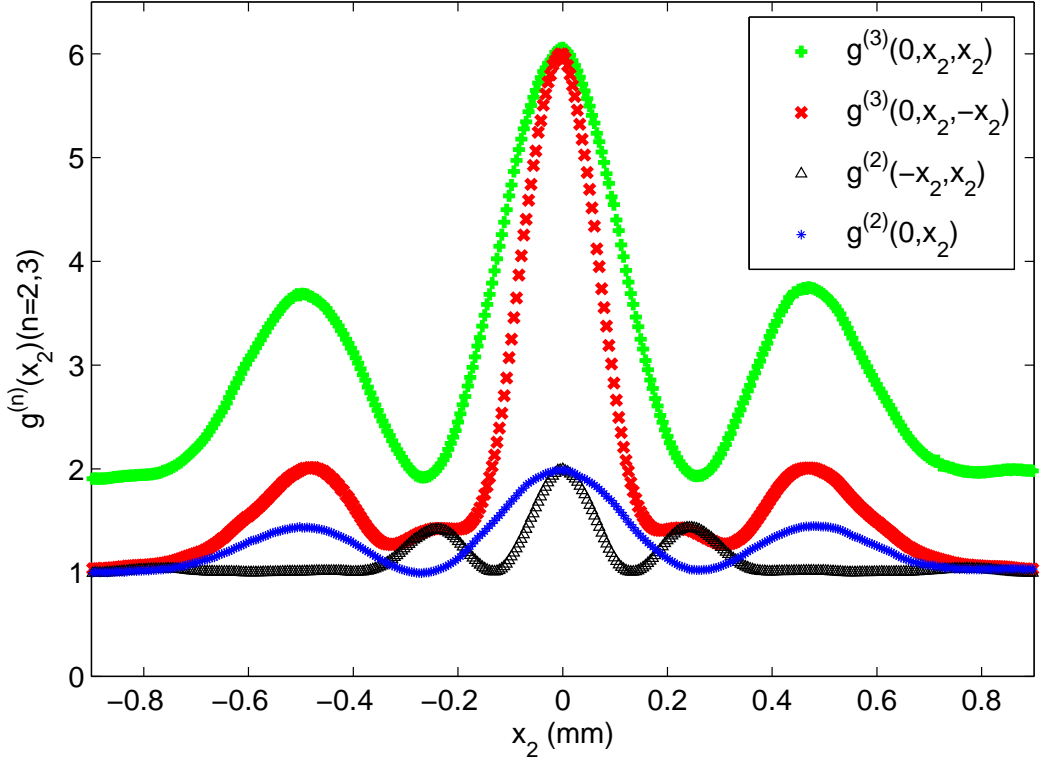}}\caption{(Color online)
Interference fringes of 2nd and 3rd order correlation functions. Blue
asterisks: 2nd order; black triangles: 2nd order sub-wavelength fringes; green
crosses: 3rd order; red crosses: 3rd order, obtained by scanning the two
reference detectors in opposite directions $x_3=-x_2$ with fixed $x_1$=0.}
\label{inter}
\end{figure}

An outline of the experimental set-up is shown in Fig.~\ref{scheme1}. A He-Ne
laser beam of wavelength $\lambda$=632.8~nm and beam diameter $D$=3.5~mm is
projected onto a rotating ground-glass plate G to produce pseudo-thermal light.
The scattered beam is first separated by a 50:50 non-polarizing beamsplitter
(BS$_1$) into two beams. The reflected beam passes through a 50:50 BS$_2$, so
the pseudo-thermal beam is divided into altogether three beams, which are
detected by the charged-coupled-device (CCD) cameras CCD$_1$, CCD$_2$ and
CCD$_3 ($Imaging Source DMK 41BU02), respectively. A double-slit with slits of
width 200 $\mu$m and separation 400 $\mu$m is placed between G and BS$_1$. The
distances between the double-slit and three CCDs are $z_1$, $z_2$ and $z_3$,
and $z_1$=$z_2$=$z_3$=336~mm. The CCDs are operated in the trigger mode and
synchronized by the same trigger pulse. The data are saved through a USB cable
to a computer. All the experimental results were obtained by averaging about
40000 exposure frames. It should be noted that we place the double-slit before
BS$_1$ unlike in the traditional GI scheme, but it will be seen that the
results are the same as those derived for higher order correlation GI in Ref.
~\cite{Liu}.

In order to compare the visibility enhancement, we first calculate the
second-order interference with the data acquired from CCD$_1$ and CCD$_2$
according to Eq.~\ref{g2}. The experimental results are shown in
Fig.~\ref{inter}, in which the blue asterisks and black triangles denote,
respectively, the second-order interference and sub-wavelength interference
patterns~\cite{Strekalov,wang1}. Their measured visibilities are 0.328 and
0.331, respectively, which is in accordance with the theoretical value of 1/3,
within experimental error. The third-order interference patterns are calculated
according to Eq.~\ref{g3} with the data from the three CCDs. The green cross
curve is obtained by scanning the two reference detectors synchronously,
namely, keeping $x_3=x_2$ with the position of CCD$_1$ fixed at $x_1$=0. When
the two reference detectors are scanned in opposite directions, namely,
$x_3=-x_2$ with $x_1$=0, we obtain the red cross curve. The measured
visibilities are 0.49 and 0.70, which are much higher than those of the
second-order case.

\begin{figure}[b]
\centerline{\includegraphics[width=14cm]{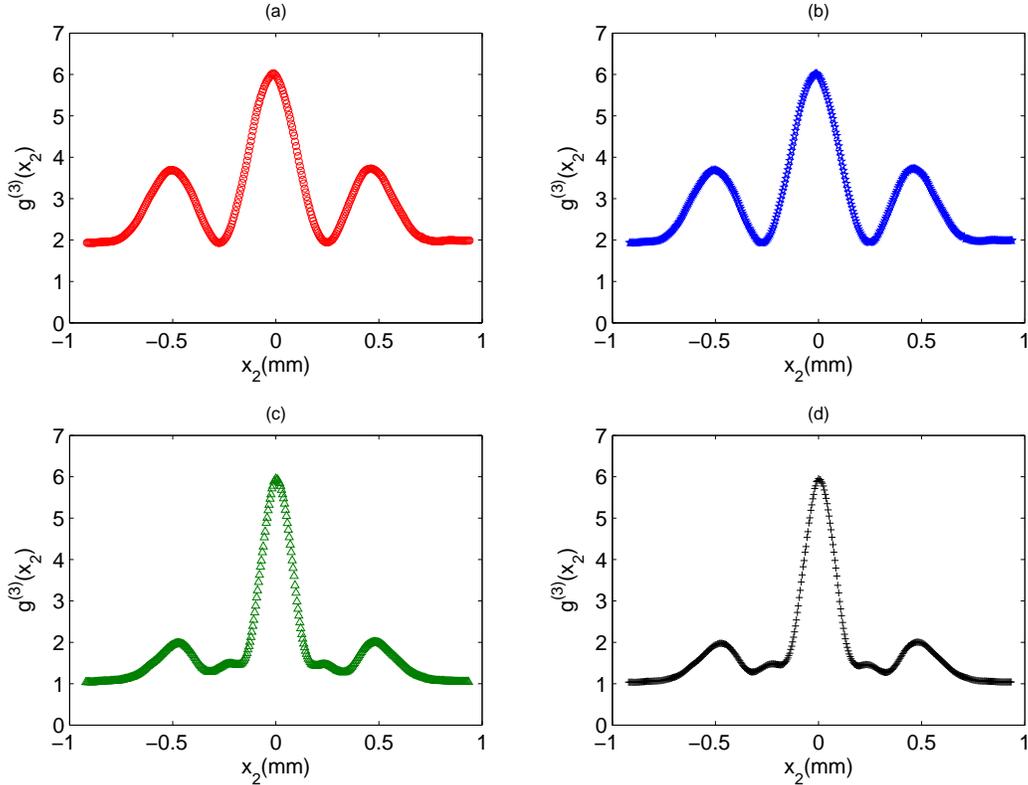}}\caption{(Color online)
Experimental results of third-order double-slit interference. (a), (c) obtained according to Eq.~$\ref{g3}$ ; (b), (d) according to Eq.~$\ref{sumg3}$; (a),(b) obtained when keeping $x_2=x_3$ with $x_1=0$; (c), (d) keeping $x_2=-x_3$.} \label{g3zh}
\end{figure}
According to Eq.~\ref{sumg3}, we can see that the visibility of third-order
interference when $x_3=x_2$ is due to the contribution from the sum of $\Delta
g^{(2)}_{12}(x_1,x_2)$, $\Delta g^{(2)}_{13}(x_1,x_3)$ and $\Delta
g^{(3)}_{123}(x_1,x_2,x_3)$, while when $x_3=-x_2$ we have to include the extra
term $\Delta g^{(2)}_{23}(x_2,x_3)$. It is precisely the contribution of this
factor which makes the latter visibility higher. It is found that, where the
secondary peaks of the sub-wavelength interference pattern occur (black
triangles), there is a pair of smaller and narrower interference peaks at the
corresponding positions in the red cross curve, while there is none in the
green cross one, which is the result of the superposition of $\Delta
g^{(2)}_{23}(x_2,x_3)$ and $\Delta g^{(3)}_{123}(x_1,x_2,x_3)$.

The two curves of the third-order interference patterns shown in Figs.~\ref{g3zh} (a), (b)and  Figs.~\ref{g3zh} (c), (d)) are processed and plotted according to Eqs.~\ref{g3} and ~\ref{sumg3}, respectively, with the condition of $x_3=x_2$($x_3=-x_2$). We can see that they are almost the same. This experimentally proves not only the equivalence of Eqs.~\ref{g3} and ~\ref{sumg3} and but also that the intensity fluctuation correlation of thermal light is the reason that leads to the double-slit and ghost imaging phenomena.

\begin{figure}[t]
\centerline{\includegraphics[width=14cm]{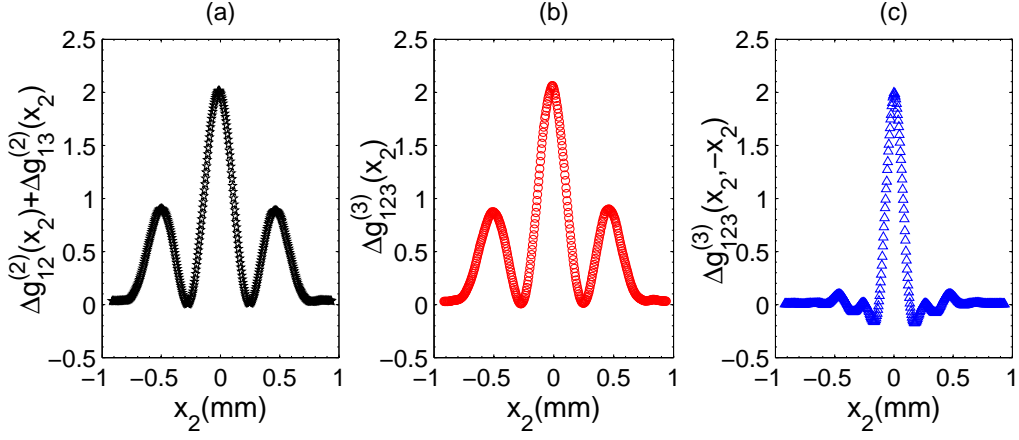}}\caption{(Color online)
Comparison of $\Delta g^{(2)}_{12}(x_2)+\Delta g^{(2)}_{13}(x_2)$, $\Delta g^{(3)}_{123}(x_2)$ and $\Delta g^{(3)}_{123}(x_2,-x_2)$.} \label{deltg123}
\end{figure}

We have also calculated the third-order intensity fluctuation correlation $\Delta g^{(3)}_{123}$ and the sum of two second-order intensity fluctuation correlations $\Delta g^{(2)}_{12}+\Delta g^{(2)}_{13}$ according to Eqs. ~\ref{deltg2} and ~\ref{deltg3}, which are shown in Fig.~\ref{deltg123}. It can be seen in Figs.~\ref{deltg123}(a) and (b) that the third-order intensity fluctuation contribution
to the final interference is approximately equal to the sum of the two second-order intensity fluctuation correlations only when the condition of synchronously scanning the reference detectors is satisfied, which is also in good agreement with Ref.~\cite{Liu}. However, there are some differences between the curve of $\Delta g^{(3)}_{123}$ in Fig.~\ref{deltg123}(c) and that of $\Delta g^{(2)}_{12}+\Delta g^{(2)}_{13}$ in Fig.~\ref{deltg123}(a) as we can see that, when the reference detectors are scanned in opposite directions $x_3=-x_2$, compared with $\Delta g^{(2)}_{12}+\Delta g^{(2)}_{13}$, the peak of $\Delta g^{(3)}_{123}$  is much narrower. Furthermore, there are two small, narrow peaks on either side of the main peak; the inner, relatively smaller peaks are due to the second-order intensity fluctuation correlation of the two reference beams, while the outer two larger peaks are due to the second-order correlations of beam 1 with beams 2 and 3. It is interesting that there are some negative values in the curve of Fig.~\ref{deltg123} (c). This is due to interference from the interaction terms when $x_3=-x_2$, as $\Delta g^{(3)}_{123}$ in Eq.~\ref{deltg3} can be negative.

In conclusion, we have experimentally realized third-order double-slit
interference with pseudo-thermal light in the high-intensity limit by recording
the intensities in three separate optical paths. The experimental results show
that the visibility of third-order interference is much higher than that of the
second-order case, and is much higher when two reference detectors are scanned
in opposite directions than when in the same direction, due to the contribution
of the cross-correlation fluctuation $\Delta g^{(2)}_{23}$. Through comparison
of $\Delta g^{(3)}_{123}$ and $\Delta g^{(2)}_{12}+\Delta g^{(2)}_{13}$, we
have also experimentally verified the theoretical prediction obtained in
ref.~\cite{Liu} that the enhanced visibility of $N$th-order double-slit
interference is a consequence of the contribution of $N$th-order intensity
fluctuation correlations, which is equal to the sum of the contributions from
all the $(N-1)$th-order correlations. Although in this experiment the
double-slit is placed before the beamsplitter, unlike in traditional GI, this
conclusion is the same as that derived for higher order correlation GI in Ref.
~\cite{Liu}.

This work was supported by Grant Nos 11204117 and 11005055 of the National Natural Science Foundation of China, Grant No. 2010CB922904 of the National Basic Research Program of China, Grant No. 2011AA20102 of the Hi-Tech Research and Development Program of China, Grant Nos. L2012001 and LJQ2011005 of the Education Department of Liaoning Province, and Grant No 20111034 of the Ph. D. Program Foundation of Liaoning Province.

\end{document}